\documentstyle[preprint,aps]{revtex}
\begin{document}
\draft
\preprint{CU-TP-862}
\title{Nuclear dependence in Drell-Yan transverse momentum 
distribution}
\author{Xiaofeng Guo}
\address{Department of Physics, Columbia University \\
         New York, NY 10027, USA}
\date{November 24, 1997}
\maketitle

\begin{abstract}

We calculate the nuclear enhancement in Drell-Yan  transverse
momentum distribution  in hadron-nucleus collisions.  In terms 
of multiple scattering picture, we compute the contributions from
double scattering. 
Applying the generalized factorization theorem, we express the nuclear
enhancement in terms of twist-4 nuclear parton correlation functions.
We demonstrate that these twist-4 nuclear parton correlation functions  
are universal, and are the same as those used to explain the nuclear 
dependence in di-jet 
momentum imbalance and in direct photon production. 
 Using the known information on the twist-4 parton correlation 
functions,  we estimate the nuclear enhancement for 
Drell-Yan pair in large $q_T$ region. We also discuss the 
source of nuclear suppression in small $q_T$ region. 
\end{abstract} 
\vspace{0.2in}
\pacs{11.80.La, 12.38.Bx, 13.85.Qk, 24.80.-x}

\section{Introduction}
\label{sec:intro}

Recently, it is found in experiment that 
the transverse momentum  distribution of
Drell-Yan pair  show anomalous nuclear dependence \cite{NA10}. The observed
nuclear enhancement in average squared 
transverse momentum, $\langle q_T^2 \rangle $, grows approximately as 
$A^{1/3}$ with $A$ the atomic number of the nuclear target 
\cite{NA10,E772}. 

Since the invariant mass of the observed Drell-Yan pairs, $Q^2$, is
large, it is necessary to understand the nuclear enhancement at partonic
level. Any single scattering with a large $Q^2$ is localized in space, and
hence, does not introduce the $A^{1/3}$ type nuclear dependence. The fact that
the observed nuclear enhancement grows with nuclear size clearly indicate 
that the anomalous nuclear enhancement should be a result of multiple
scattering of a parton inside the nucleus. 

In terms of factorization at higher twist, Luo, Qiu and Sterman (LQS) have 
developed a consistent treatment of multiple scattering at partonic level
\cite{LQS}. LQS expressed the nuclear dependence of di-jet momentum imbalance
in photo-nucleus collisions in terms of twist-4 nuclear parton
distributions. Using the Fermi Lab E683 data, LQS estimated the size of the
relevant twist-4 parton distributions to be of the order of $0.05-0.1$ GeV$^2$
times typical twist-2 parton distributions \cite{LQS2}.

The transverse momentum distribution for Drell-Yan process, 
$d\sigma / dQ^2 dq_T^2$,  is a typical
two-scale problem in perturbative QCD. 
When $q_T^2$ is the same order as $Q^2$, $d\sigma/dQ^2\,dq_T^2$
becomes an one scale process, and the conventional perturbative QCD
calculation works well.  However, when $q_T^2$ is much less than
$Q^2$, $d\sigma/dQ^2\,dq_T^2$ becomes perturbatively unstable, and
is proportional to $1/q_T^2$ as $q_T^2 \rightarrow 0$.  In addition,
the conventionally calculated partonic parts develop two powers of
large logarithm, $\log^2(Q^2/q_T^2)$, for every power of $\alpha_s$,
due to collinear and soft gluon radiation from incoming beam partons.
It is therefore necessary to perform the resummation of these large
logarithms in order to get a correct Drell-Yan $q_T^2$ spectrum for
the region when $q_T^2 \ll Q^2$ \cite{CSS}.

In terms of multiple scattering picture, nuclear dependence of Drell
Yan $q_T^2$ spectrum is a result of multiple scattering between
incoming beam parton and nuclear matter before the Drell-Yan lepton
pair is produced.  When $Q^2$ and $q_T^2$ are both large, double
scattering should be dominant.  Method developed by LQS \cite{LQS}
can be naturally applied for calculating nuclear dependence of
Drell-Yan $d\sigma/dQ^2\,dq_T^2$.  However, when $Q^2$ is large and 
$q_T^2$ is relatively small, calculating the nuclear dependence of
Drell-Yan $d\sigma/dQ^2\,dq_T^2$ will face the same complication as
calculating $d\sigma/dQ^2\,dq_T^2$ itself.  As emphasized in last
paragraph, when $q_T^2 \ll Q^2$, Drell-Yan $d\sigma/dQ^2\,dq_T^2$ is
not perturbatively stable without resummation of large logarithms.
Due to possible interference of multiple scattering and collinear and
soft radiation, the resummation of large logarithms may not be as
straightforward \cite{GQS}.  

Because of two physical scales, $Q^2$ and $q_T^2$, even at lowest order, 
there are two types of double scattering: soft-hard double 
scattering and double-hard double scattering. In soft-hard process, 
the parton from the beam first absorbs a soft gluon at the amplitude 
level, and then undergoes 
a hard scattering with another parton from the nucleus to produce the 
virtual photon $\gamma ^*$ at $Q^2$ and $q_T^2$. In double-hard process, 
the parton from the beam  first undergoes a hard
scattering with a parton from the nucleus to produce a quark $q$ (or an 
anti-quark) at the order of $q_T^2$, then the produced quark $q$ (or 
anti-quark) 
annihilates with an anti-quark (or a quark) to produce the virtual photon 
at $Q^2$ and $q_T^2$.
As explained later, the double-hard process resembles  the classical 
double scattering picture, while 
the soft-hard process does not. These two types of processes have 
opposite sign
at the amplitude level. When $q_T \rightarrow 0$, the amplitude of 
soft-hard and double-hard scattering cancel each other. 
Both the soft-hard and double hard
process contribute to nuclear enhancement, but their interference gives
suppression effect. In large $q_T$ region, these two processes require 
very different parton flux, and the interference term is less
important. However, in smaller $q_T$ region, the interference term becomes 
more important and 
can reduce the enhancement, or even give suppression effect.

In this paper, we calculate the nuclear dependence of  Drell-Yan 
$d\sigma/dQ^2\,dq_T^2$ for large $q_T$ region, where fixed order 
perturbative calculation works well. Nuclear dependence in small $q_T$ 
region will be addressed in Ref. \cite{GQS}. This paper is organized as the
following, in Sec.II, we outline the general
formalism and consideration for double scattering contribution; in Sec.III, 
we compute the soft-hard double scattering contributions; in Sec.IV, we 
derive
the formula for double-hard scattering process;  in Sec V, we discuss the
relation between soft-hard and double-hard processes;  in Sec. VI, we present 
the numerical result and conclusions.

\section{general formalism}
\label{sec:formal}

To study the nuclear dependence of transverse momentum distribution, we
consider the differential cross section for Drell-Yan process:
\begin{equation}
  h(p') + A(p) \rightarrow \ell^+\ell^-(q) + X \ ,
\label{e1}
\end{equation}
where $p'$ is the momentum for the incoming beam hadron, and $p$ is the 
momentum per
nucleon for the nuclear target. In Eq.~(\ref{e1}), $q$ is the 
four-momentum for the virtual
photon, which decays into the lepton pair.  
In terms of contributions from multiple scattering, we expand the
differential cross section as
\begin{equation}
\frac{d\sigma_{hA}}{dQ^2 dq_T^2 dy}
  = \frac{d\sigma_{hA}^S}{dQ^2 dq_T^2 dy} 
   +\frac{d\sigma_{hA}^D}{dQ^2 dq_T^2 dy}
   + ... \ .
\label{e2}
\end{equation}
The superscript ``S'' and ``D'' denote contributions from the ``single 
scattering'' and ``double scattering'', respectively, and ``...''
represents contributions from even higher multiple scattering. 
In Eq.~(\ref{e2}), $Q$ is the invariant mass of the lepton 
pair, $q_T$ is the total transverse momentum of
the lepton pair, and  $y$ is its rapidity 
\begin{equation}
y=\frac{1}{2} \,{\rm ln}\, \frac{q_0+q_z}{q_0-q_z} \ .
\label{y}
\end{equation}
As discussed above, single scattering  
is localized, and therefore, $d\sigma_{hA}^S/dQ^2 dq_T^2 dy$ does not 
contribute to anomalous nuclear dependence. It is 
proportional to $A$, modular nuclear shadowing effect.  On the other 
hand, clear $A^{1/3}$ dependence shown in the experimental 
data \cite{E772} indicates that multiple 
scattering (higher than double scattering) is less important in 
proton-nucleus collisions. 
In this paper, we compute the double scattering
contribution $d\sigma_{hA}^D/dQ^2 dq_T^2 dy$. 

For fix target experiments, often the variable $x_F=2q_z/\sqrt{s}$ is used. 
We can easily 
convert our results $d\sigma_{hA}/dQ^2 dq_T^2 dy$ to 
$d\sigma_{hA}/dQ^2 dq_T^2 dx_F$ by using the relation 
\begin{equation}
dQ^2dq_T^2dy=
\frac{1}{\sqrt{x_F^2+4 (Q^2+q_T^2)/s}} \, dQ^2dq_T^2dx_F\ .
\label{y-xf}
\end{equation}
In order to compare with experimental data, we calculate the
ratio of total differential cross section  and
single scattering contribution, 
\begin{equation}
R=\frac{d\sigma_{hA}}{dQ^2 dq_T^2 dy} 
  \left/  \frac{d\sigma_{hA}^S}{dQ^2 dq_T^2 dy} \right. \,
 \approx 1+\frac{d\sigma_{hA}^D}{dQ^2 dq_T^2 dy} 
  \left/  \frac{d\sigma_{hA}^S}{dQ^2 dq_T^2 dy} \right. \, .
\label{e3}
\end{equation}

For single scattering, Fig.~\ref{fig1} shows the lowest 
order Feynman diagrams 
that contribute to transverse momentum
distribution of Drell-Yan pair. According to factorization 
theorem \cite{Factorization}, the single
scattering contribution can be expressed as
\begin{eqnarray}
\frac{d\sigma^S}{dQ^2dq_T^2dy} &=&
\frac{2\alpha_{em}^2 \alpha_s}{3Q^2}\, \sum_{a,c} \,
\frac{1}{2s }
\int \frac{dx'}{x'} \frac{dx}{x}  \, 
f_{c/h}(x')\,  f_{a/A}(x) \,
|\overline{M}_{ca\rightarrow \gamma^*}|^2
\nonumber \\ 
&\ & {\hskip 0.8in} \times 
\frac{1}{x's+u-Q^2}
\delta \left( x+\frac{x'(t-Q^2)+Q^2}{x's+u-Q^2} \right) \ .
\label{single}
\end{eqnarray}
In Eq.~(\ref{single}), $x'$ is the momentum fraction of parton 
``c'' from the beam
hadron $h$, and $x$ is the momentum fraction of parton 
``a'' from the the nuclear
target. $f_{c/h}(x')$ and $f_{a/A}(x)$ are the parton distribution functions
for the hadron and the nucleus respectively. $s, t, u$ are the invariant
variables defines as
\begin{mathletters}
\label{stu} 
\begin{eqnarray}
s &=& (p+p')^2 =2p\cdot p'\, ; 
\label{s} 
\\  
t &=& (p'-q)^2 =-2p' \cdot q+Q^2\, ; 
\label{t} 
\\  
u &=& (p-q)^2 =-2p\cdot q+Q^2\ .
\label{u}
\end{eqnarray}
\end{mathletters} 
The averaged matrix element square 
$|\overline{M}_{ab\rightarrow \gamma^*}|^2$ can
be computed from the Feynman diagrams shown in Fig.~\ref{fig1} 
\cite{Fields}.
\begin{mathletters}
\label{sm2}
For the annihilation diagram 
\begin{eqnarray}
|\overline{M}_{q\bar{q}\rightarrow \gamma^*}|^2 &=& 
\frac{4}{9} e_q^2 \cdot 2\,
\left( \frac{\hat{t}}{\hat{u}} +\frac{\hat{u}}{\hat{t}} \right.
\left. +\frac{2Q^2\hat{s}}{\hat{t}\hat{u}} \right) \ ;
\label{sm2a}
\end{eqnarray}
for ``Compton'' diagram with the quark from the beam and the gluon from the
target 
\begin{eqnarray}
|\overline{M}_{qg\rightarrow \gamma^*}|^2 &=& 
\frac{1}{6} e_q^2 \cdot 2\, 
\left( \frac{-\hat{t}}{\hat{s}} +\frac{\hat{s}}{-\hat{t}} \right.
\left. -\frac{2Q^2\hat{u}}{\hat{t}\hat{s}} \right) \ ;
\label{sm2b}
\end{eqnarray}
for ``Compton'' diagram with the gluon from the beam and the quark from the
target 
\begin{eqnarray}
|\overline{M}_{gq\rightarrow \gamma^*}|^2 &=& 
\frac{1}{6}  e_q^2 \cdot 2\, 
\left( \frac{\hat{s}}{-\hat{u}} +\frac{-\hat{u}}{\hat{s}} \right.
\left. -\frac{2Q^2\hat{t}}{\hat{s}\hat{u}} \right) \ .
\label{sm2c}
\end{eqnarray}
\end{mathletters}
In Eq.~(\ref{sm2}), the factor of coupling constants $e^4g_s^2$ is 
already included in the over 
all factor in Eq.~(\ref{single}). $\hat{s}, \hat{t}, \hat{u}$ are defined as:
\begin{mathletters}
\label{stu1}
\begin{eqnarray}
\hat{s} &=& (xp+x'p')^2 = xx's\, ; 
\label{s1}  \\ 
\hat{t} &=& (x'p'-q)^2  =x'(t-Q^2)+Q^2\, ; 
\label{t1}  \\
\hat{u} &=& (xp-q)^2  =x(u-Q^2)+Q^2\ .
\label{u1}
\end{eqnarray}
\end{mathletters}

According to generalized factorization theorem \cite{QS},
double scattering contribution to the cross section can be factorized 
into the following form: 
\begin{equation}
 \frac{d\sigma_{hA \rightarrow l\bar{l}}^D}{dQ^2 dq_T^2 dy}
= \sum_c \int dx' f_{c/h}(x') \, 
\frac{d\hat{\sigma}_{cA \rightarrow l\bar{l}}^D}{dQ^2 dq_T^2 dy} \, , 
\label{e4}
\end{equation}
where $d\hat{\sigma}_{cA}^D$ represents the double scattering contribution
between a beam parton of flavor ``c'' and the nuclear target.  The
$f_{c/h}(x')$ is the normal twist-2 parton distribution of flavor
``c'' within a hadron ``h''. 
 We can simplify the calculation by 
integrating over the leptonic part of phase space first.
\begin{equation}
\frac{d\hat{\sigma}_{cA \rightarrow l \bar{l}}^D}{dQ^2dq^2_Tdy}
=\frac{1}{2\pi}\frac{e^2}{3Q^2} 
\frac{d\hat{\sigma}_{aA \rightarrow \gamma^{\ast}}}{dQ^2dq^2_Tdy} \ .
\label{e4b}
\end{equation}
According to the  factorization at higher twist, 
$d\hat{\sigma}_{cA}^D /dQ^2 dq_T^2 dy$ can be expressed in the following
factorized form:
\begin{equation}
\frac{d\hat{\sigma}_{cA}^D}{dQ^2 dq_T^2 dy} =\sum_{{i}}
\int dx dx_1 dx_2 \, T_{{i}}(x,x_1,x_2,p) 
 H_{{i}}(x,x_1,x_2,p,p',q).
\label{e5a}
\end{equation}
The graphical representation of Eq.~(\ref{e5a}) is shown in Fig.~\ref{figa}.

At lowest order, there are two types of double scattering. Fig.~\ref{fig2}
shows sample diagrams at amplitude level 
for these two types of double scattering. 
In double scattering process, there are two partons ``a'' and ``b'' 
from the nucleus participate in the scattering process. The kinematics 
can only fix one parton momentum. We need to integrate over the other 
parton's momentum. As shown in  Fig.~\ref{fig2}, because of the extra
scattering, we have two propagators, that is, two possible poles. 
Therefore, the leading contribution of the integration    
over the extra parton momentum is given by the residues at these poles.  
At fixed $Q$ and $q_T$, we cannot have both of these two poles at the 
same time. The total double scattering contribution is the sum of the 
residue contributions from 
these two different poles. Correspondingly, there are two 
types of double scattering. Taking the residue corresponding to the pole of 
the first propagator, as shown in Fig.~\ref{fig2}a,  
is equivalent to say that the first parton (gluon) is soft. We call it
soft-hard double scattering.   In this kind of
processes, the parton ``c'' from the beam first absorbs a soft gluon 
at the amplitude level, and then undergoes a hard scattering with 
another parton from 
the nucleus to produce the virtual photon 
$\gamma ^*$. The large momentum transfer occurs only in the hard
scattering. 
Absorption of the gluon modifies the effective parton flux, so as 
the overall cross section.
Taking the residue from the pole of  the second propagator, as shown 
in Fig.~\ref{fig2}b, correspond to a situation when both partons 
$a$ and $b$ have non-zero momenta. We call this kind of processes 
double-hard double scattering. 
 In this type of processes, the parton ``c'' first have a hard
scattering with a parton ``b'' from the nucleus to produce a 
quark $q$ (or an 
anti-quark) at a scale of $q_T^2$, then the produced quark $q$
annihilate with an anti-quark (or a quark) ``a'' to produce the 
virtual photon at a scale of $Q^2$. More discussions about the 
difference and the 
relations of soft-hard and double-hard processes is given in Sec. V. 
In the following two sections, we derive the contributions from
soft-hard and double-hard processes respectively.
  
\section{soft-hard double scattering contribution}
\label{sec:soft-hard}

In this section, we derive the contribution from soft-hard scattering. At
lowest order, the soft-hard double scattering has three types of 
subprocesses,
as show in Fig.~\ref{fig3}, Fig.~\ref{fig5} and Fig.~\ref{fig6}. 
These three types of subprocesses correspond to
adding two soft gluons to the lowest order ``Annihilation'' and ``Compton''
subprocesses. In the following subsections, we present the derivations for one
subprocess, and provide the results for other subprocesses.

\subsection{Factorization and $k_T$ expansion for soft-hard processes}
\label{subsec:sh1}

Consider subprocess shown in Fig.~\ref{fig3}. There are three independent
four-momentum  linking the partonic part and corresponding 
two-quark-two-gluon matrix element. In the center of mass frame of high energy
collision, all partons inside the nucleus are moving almost parallel to each 
other, along the momentum of the nucleus. All three parton momenta can be
approximately replaced by the components collinear to the hadron
momentum. After such collinear expansion, the double scattering contribution
for contribution from process shown in Fig.~\ref{fig3}a can be written as:
\begin{equation}
\frac{d\hat{\sigma}^{(SH)}_{qA\rightarrow\gamma ^*}}{dQ^2dq^2_Tdy} 
= \frac{1}{2x's}\,\int dx\, dx_{1}\, dx_{2}\,
\int d^{2}k_{T}\, \overline{T}(x,x_{1},x_{2},k_{T},p)\,
\overline{H}(x'p',x,x_{1},x_{2},k_{T},p,q)\ ,
\label{e5}
\end{equation}
where $2x's$ is the flux factor between the incoming beam quark
and the nucleus, and $x'p'$ is the momentum carried by the beam quark. 
We keep the $k_T$ for the soft gluons  in order to extract a double scattering
contribution beyond the leading twist. The superscript ``$SH$'' denotes 
``soft-hard'' double scattering. In Eq.~(\ref{e5}), the 
two-quark-two-gluon matrix element, $\overline{T}$, is defined as
\begin{eqnarray}
\overline{T}(x,x_{1},x_{2},k_{T},p) 
&=& \int \frac{dy^{-}}{2\pi}
     \frac{dy_{1}^{-}}{2\pi} \frac{dy_{2}^{-}}{2\pi} 
     \frac{d^{2}y_{T}}{(2\pi)^2} \nonumber \\
&\ & \times e^{ixp^{+}y^{-}}\, e^{ix_{1}p^{+}y_{1}^{-}}\,
            e^{-i(x_{1}-x_{2})p^{+}y_{2}^{-}}\,
            e^{-ik_{T}\cdot y_{T}} \nonumber \\
&\ & \times \frac{1}{2}\langle p_{A} | 
            A^+(y_{2}^{-},0_{T})\, \bar{\psi}_q(0)\, \gamma^+ \,
            \psi_q(y^{-})\, A^+(y_{1}^{-},y_{1T}) | p_{A}\rangle\ .
\label{e6}
\end{eqnarray}
The corresponding partonic part $\overline{H}$ is given by the
diagrams shown in Fig.~\ref{fig3}a, with gluon lines contracted with
$p^{\rho}p^{\sigma}$, quark lines from the target traced with
$(\gamma\cdot p)/2$, and quark lines from the beam traced with 
$(\gamma\cdot (x'p'))/2$.  Here, we work in Feynman gauge, 
in which the leading contribution from the gluon field operators is
$A^{\rho}\approx A^+(p^{\rho}/p^+)$.

In order to extract the lowest order high-twist contribution, we expand 
the partonic part
 $\overline{H}$ introduced in
Eq.~(\ref{e5}) at $k_T=0$, 
\begin{eqnarray}
\overline{H}(x'p',x,x_{1},x_{2},k_{T},p,q) 
&=& \overline{H}(x'p',x,x_{1},x_{2},k_{T}=0,p,q) \nonumber \\
&+& \left. \frac{\partial \overline{H}}{\partial k_{T}^{\alpha}}
    \right|_{k_{T}=0}\ k_{T}^{\alpha}\, 
+\, \left. \frac{1}{2}\, \frac{\partial^{2}\overline{H}}
                {\partial k_{T}^{\alpha} \partial  k_{T}^{\beta}} 
    \right|_{k_{T}=0}\ k_{T}^{\alpha}\, k_{T}^{\beta} + \ldots\ .
\label{e7}
\end{eqnarray}
In the right-hand-side of Eq.~(\ref{e7}), the first term is the
leading twist eikonal contribution, which does not correspond to
physical double scattering, but simply makes the single-scattering
matrix element gauge invariant.
The second term vanishes after integrating over $k_T$.  The
third term will give the finite contribution to the double 
scattering process. 
Substituting Eq.~(\ref{e7}) into
Eq.~(\ref{e5}), and integrating over $d^2k_T$, we obtain
\begin{eqnarray}
\frac{d\hat{\sigma}^{(SH)}_{qA\rightarrow\gamma^*}}{dQ^2dq^2_Tdy} 
&=& \frac{1}{2x's}\, 
    \int \frac{dy^{-}}{2\pi}\, \frac{dy_{1}^{-}}{2\pi}\,  
      \frac{dy_{2}^{-}}{2\pi}\,
\frac{1}{2}\, 
     \langle p_{A} |  F_{\alpha}^{\ +}(y_{2}^{-})\, \bar{\psi}_q(0)\,
           \gamma^+\, \psi_q(y^{-})\, F^{+\alpha}(y_1^{-}) 
     | p_{A}\rangle \nonumber \\
&\times &
    \left(-\frac{1}{2}g^{\alpha\beta}\right)
\left[\, \frac{1}{2}\, \frac{\partial^{2}}
                   {\partial k_{T}^{\alpha} \partial k_{T}^{\beta}}\,
    H(y^-,y_1^-,y^-_2,k_T=0,p,q)\, \right]\ .
\label{e10}
\end{eqnarray}
In Eq.~(\ref{e10}), the modified partonic part $H$ is defined as
\begin{eqnarray}
H(y^-,y_1^-,y^-_2,k_T,p,q) 
&=& \int dx\, dx_{1}\, dx_{2}\, 
     e^{ixp^{+}y^{-}}\, e^{ix_{1}p^{+}y_1^{-}}\,
     e^{-i(x_{1}-x_{2})p^{+}y_{2}^{-}}  \nonumber \\
&\ & \times \overline{H}(x'p',x,x_1,x_{2},k_T,p,q)\ ,
\label{e11}
\end{eqnarray}
where the partonic part $\overline{H}$ is defined by Eq.~(\ref{e5}), 
and is given by diagrams shown in
Fig.~\ref{fig3}. 
In Eq.~(\ref{e10}), $F^{+\alpha}=F^{\beta\alpha}n_{\beta}$, and
$F^{\beta\alpha}$ is the field strength, and vector
$n_{\beta}=\delta_{\beta +}$.
In obtaining Eq.~(\ref{e10}), we use the factor $k_T^{\alpha} k_T^{\beta}$
in Eq.~(\ref{e7}) to convert the field $A^+$ into field strength by 
partial integration. 

\subsection{Integration over the parton momentum fractions}
\label{subsec:sh2} 

From Eqs.~(\ref{e10}) and (\ref{e11}), all integrals
of $x, x_1$ and $x_2$  can now be done explicitly without knowing the
details of the multi-parton matrix elements.

Consider the diagram shown in Fig.~\ref{fig3}a.  The final state
photon-gluon two particle phase space can be written as
\begin{equation}
\Gamma^{(2)}=\frac{1}{16\pi^2}\, \frac{1}{x's+u-Q^2}\,
\delta\left(x+x_1+\frac{\hat{t}-k_T^2-2k_T\cdot q}{x's+u-Q^2}\right)\ .
\label{e12}
\end{equation}
In deriving Eq.~(\ref{e12}), we have omitted the factor $dQ^2dq^2_Tdy$, 
due to the definition of the invariant cross section. 
$k_T^2=-k_{T\alpha}k_T^{\alpha}$, and $\hat{t}$ is defined in Eq.~(\ref{t1}). 

Using Eq.~(\ref{e12}), the contribution to $\overline{H}$ from the diagram
shown in Fig.~\ref{fig3}a can be expressed as
\begin{eqnarray}
\overline{H}_{1a} &=& 
\frac{\alpha_s}{4\pi}\, C_{1}\, \frac{1}{x's+u-Q^2}\,
      \hat{H}_{1a}(x,x_1,x_{2}) \nonumber \\
&\times &
\frac{1}{x_1-x_{2}-\frac{k_T^2}{x's}-i\epsilon}\,
\frac{1}{x_1-\frac{k_T^2}{x's}+i\epsilon}\,
\delta\left(x+x_1+\frac{\hat{t}-k_T^2-2k_T\cdot q}{x's+u-Q^2}\right)\ ,
\label{e13}
\end{eqnarray}
where the subscript ``$1a$'' has following convention: ``$1$''
stands for the first type subprocess, shown in Fig.~\ref{fig3}; ``$a$''
corresponds to diagram in
Fig.~\ref{fig3}a.  In Eq.~(\ref{e13}), the factor $C_1$ is an overall
color factor for first type  subprocess. 
The function
$\hat{H}_{1a}$ in Eq.~(\ref{e13}) is 
given by  
\begin{equation}
\hat{H}_{1a}=\frac{1}{4}\, \frac{1}{x's}\,
{\rm Tr}\left[\gamma\cdot(x'p'+k_T)\gamma\cdot p\, \gamma\cdot(x'p'+k_T)
         R_{1a}^{\beta\nu}\gamma\cdot p\, L_{1a}^{\alpha\mu}\right]
\left(-g_{\alpha\beta}\right)\left(-g_{\mu\nu}\right)\ ,
\label{e14}
\end{equation}
where $R_{1a}^{\beta\nu}$ and $L_{1a}^{\alpha\mu}$ are the right and
left blob, respectively, as shown in Fig.~\ref{fig3}a.  These blobs
include all possible tree Feynman diagrams with the external partons
shown in the figure.  Substituting Eq.~(\ref{e13}) into Eq.~(\ref{e11}), 
we obtain 
\begin{eqnarray}
H_{1a}&=&
\frac{\alpha_s}{4\pi}\, C_{1}\, \frac{1}{x's+u-Q^2}\,
\int dx_{1}\,dx_{2}\, dx\, e^{ixp^{+}y^{-}}\,e^{ix_{2}p^{+}y^{-}_2}\,
e^{ix_{1}p^{+}(y_1^{-}-y^{-}_2)}\,
\nonumber \\
&\ & \ \ \ 
\times
\frac{1}{x_1-\frac{k_T^2}{x's}+i\epsilon}\, 
\frac{1}{x_1-x_{2}-\frac{k_T^2}{x's}-i\epsilon}
\nonumber \\
&\ & \ \ \ \times 
\delta\left(x+x_1+\frac{\hat{t}-k_T^2-2k_T\cdot q}{x's+u-Q^2}\right)\,
       \hat{H}_{1a}(x,x_1,x_{2}) \ .
\label{e15}
\end{eqnarray}
After performing $dx$
by the $\delta$-function, and $dx_1$ and $dx_{2}$ by contour,  we derive
\begin{eqnarray}
H_{1a} &=&
(\pi\alpha_s)\, C_{1}\, \frac{1}{x's+u-Q^2}\,
e^{i\bar{x}p^{+}y^{-}}\, 
e^{i\left(k_T^2/x's\right)p^{+}(y^{-}_1-y^{-}_2)} \nonumber \\
&\ & \times
\theta(-y^{-}_2)\, \theta(y^{-}-y^{-}_1)\, 
\hat{H}_{1a}(\bar{x},x_1,x_{2})\ ,
\label{e16}
\end{eqnarray}
where the $\theta$-functions result from the contour integrations,
and the momentum fractions for the function $\hat{H}_{1a}$ are
defined as 
\begin{mathletters}
\label{e17}
\begin{eqnarray}
\bar{x}&=&-\frac{\hat{t}-2k_T \cdot q -k_T^2}{x's+u-Q^2} 
           -\frac{k_T^2}{x's}
\label{e17a} \\
x_1&=&\frac{k^2_T}{x's}\ ;
\label{e17b} \\
x_{2}&=&0 \ .
\label{e17c} 
\end{eqnarray}
\end{mathletters}

Similarly, we derive contribution from the diagram shown in
Fig.~\ref{fig3}b as
\begin{eqnarray}
H_{1b} &=&
(\pi\alpha_s)\, C_{1}\, \frac{1}{x's+u-Q^2}\,
e^{ixp^{+}y^{-}}\, 
e^{i\left(k_T^2/x's\right)p^{+}(y_1^{-}-y^{-}_2)} \nonumber \\
&\ & \times
\theta(y^{-}_2-y_1^{-})\, \theta(y^{-}-y^{-}_2)\, 
\hat{H}_{1b}(x,x_1,x_{2})\ ,
\label{e18}
\end{eqnarray}
where $x$ is defined as
\begin{equation}
x =-\frac{\hat{t}}{x's+u-Q^2} \, ,
\label{ex} 
\end{equation}
and $x_1$ and $x_{2}$ are defined in
Eqs.~(\ref{e17b}) and (\ref{e17c}).  
Similar to Eq.~(\ref{e14}), the partonic part 
$\hat{H}_{1b}$ is given by 
\begin{equation}
\hat{H}_{1b}=\frac{1}{4}\, 
{\rm Tr}\left[\gamma\cdot(x'p')
         R_{1b}^{\beta\nu}\gamma\cdot p\, L_{1b}^{\alpha\mu}\right]
\left(-g_{\alpha\beta}\right)\left(-g_{\mu\nu}\right)\ .
\label{e19}
\end{equation}
The diagram shown in Fig.~\ref{fig3}c has following contribution
\begin{eqnarray}
H_{1c} &=&
(\pi\alpha_s)\, C_{1}\, \frac{1}{x's+u-Q^2}\,
e^{ixp^{+}y^{-}}\, 
e^{i\left(k_T^2/x's\right)p^{+}(y_1^{-}-y^{-}_2)} \nonumber \\
&\ & \times
\theta(y_1^{-}-y^{-}_2)\, \theta(-y_1^{-})\, 
\hat{H}_{1b}(x,x_1,x_{2})\ .
\label{e20}
\end{eqnarray}
In deriving Eq.~(\ref{e20}), we used the fact that the partonic part
$\hat{H}_{1c}=\hat{H}_{1b}$ when $x_1$ and $x_{2}$ are evaluated at
the same values as listed in Eq.~(\ref{e17}).

Combining $H_{1a}$, $H_{1b}$ and $H_{1c}$ (given in
Eqs.~(\ref{e16}), (\ref{e18}), and (\ref{e20}), respectively)
together, we obtain the total contribution to $H$, defined in
Eq.~(\ref{e11}), from the first type diagrams shown in Fig.~\ref{fig3}, 
\begin{eqnarray}
H_{1} &=& H_{1a} + H_{1b} + H_{1c} \nonumber \\
&=&
(\pi\alpha_s)\, C_{1}\, \frac{1}{x's+u-Q^2}\,
e^{i\left(k_T^2/x's\right)p^{+}(y_1^{-}-y^{-}_2)}\,  
\theta(-y^{-}_2)\, \theta(y^{-}-y_1^{-})  \nonumber \\
&\ & \times
\left[ e^{i\bar{x}p^{+}y^{-}}\, 
       \hat{H}_{1a}(\bar{x},x_1,x_{2})
     - e^{ixp^{+}y^{-}}\, 
       \hat{H}_{1a}(x,x_1,x_{2}) \right] \ .
\label{e21}
\end{eqnarray}
All momentum fractions in Eqs.~(\ref{e21}) are evaluated at the values 
defined in Eqs.~(\ref{e17}) and (\ref{ex}).
In deriving Eq.~(\ref{e21}), we have dropped a term proportional to 
\[
[\,\theta(-y^{-}_2)\, \theta(y^{-}-y^{-}_1) 
-\theta(y^{-}_2-y^{-}_1)\, \theta(y^{-}-y^{-}_2) 
-\theta(y^{-}_1-y^{-}_2)\, \theta(-y^{-}_1)\,]
\longrightarrow 0\ .
\]
This is because of the phase\ exp$[ixp^+y^-]$ which effectively
restricts $y^{-}\sim 1/(xp^+) \rightarrow 0$.  Physically, it means
that all $y$ integrations in such term are localized, and therefore,
will not give any large nuclear size enhancement.

By substituting Eq.~(\ref{e21}) into Eq.~(\ref{e10}), we can obtain
the lowest order double scattering contribution from first type
diagrams shown in Fig.~\ref{fig3}. One important step in getting the
final result is taking the derivative with respect to $k_T$ as defined
in Eq.~(\ref{e10}).  Comparing Eq.~(\ref{e21}) with Eq.~(\ref{e10}),
and observing that 
\begin{equation}
\left[ e^{i\bar{x}p^{+}y^{-}}\, 
       \hat{H}_{1a}(\bar{x},x_1,x_{2})
     - e^{ixp^{+}y^{-}}\, 
       \hat{H}_{1a}(x,x_1,x_{2}) \right]_{k_T=0} = 0 \ ,
\label{e22}
\end{equation}
we found that the derivatives on the exponential\ 
exp$[i\left(k_T^2/x's\right)p^{+}(y_1^{-}-y^{-}_2)]$
do not contribute, and that we can therefore set 
exp$[i\left(k_T^2/x's\right)p^{+}(y_1^{-}-y^{-}_2)]=1$ in
Eq.~(\ref{e21}).  Substituting Eq.~(\ref{e21}) into Eq.~(\ref{e10}),
we obtain
\begin{eqnarray}
\frac{d\hat{\sigma}^{(SH)}_{qA \rightarrow \gamma^*}}{dQ^2dq^2_Tdy} 
&=&
8\pi^2\alpha_s^2\alpha_{em}\, C_1\,
\frac{1}{2x's}\, \frac{1}{x's+u-Q^2}\,
\left(-\frac{1}{2}g^{\alpha\beta}\right)\,   \nonumber \\
&\times &
\frac{1}{2}\, \frac{\partial^{2}}
                   {\partial k_{T}^{\alpha} \partial k_{T}^{\beta}}
\left[\, T_q(\bar{x},A)\, \hat{H}_{1a}(\bar{x},x_1,x_{2})
        -T_q(x,A)\, \hat{H}_{1b}(x,x_1,x_{2})\, \right] \ ,
\label{e23}
\end{eqnarray}
with $\bar{x}$, $x_1$, $x_2$ and $x$ defined in Eqs.~(\ref{e17}) and
(\ref{ex}). 
In Eq.~(\ref{e23}), $T_q(x,A)$ is the two-quark-two-gluon matrix element, 
and defined as \cite{LQS}:  
\begin{eqnarray}
T_{q}(x,A) &=& 
 \int \frac{dy^{-}}{2\pi}e^{ixp^{+}y^{-}}
 \int \frac{dy_1^{-}dy_{2}^{-}}{2\pi} 
      \theta(y^{-}-y_1^{-})\theta(-y_{2}^{-}) \nonumber \\
&\ & \times \frac{1}{2} 
     \langle p_{A}|F_{\alpha}^{\ +}(y_{2}^{-})\bar{\psi}_{q}(0)
                  \gamma^{+}\psi_{q}(y^{-})F^{+\alpha}(y_1^{-})
     |p_{A}\rangle\ .
\label{e24} 
\end{eqnarray}

\subsection{Final results for soft-hard processes}
\label{subsec:sh3}

The derivatives with respect to $k_T$ in Eq.~(\ref{e23}) are
straightforward.  It is most convenient to re-express derivatives with
respect to $k_T$ in terms of derivatives with
respect to $\bar{x}$ or $x$ \cite{LQS}.  After working out the derivatives, 
we obtain  the parton-nucleus cross section for first type soft-hard 
process   
\begin{eqnarray}
\frac{d\hat{\sigma}^{(SH)}_{1}}{dQ^2dq^2_Tdy} &=&
e_q^2\, \sigma^{DY}_0 \, (12\pi \alpha_s^2)\,
\frac{1}{2x's}\, \frac{1}{x's+u-Q^2}\,\nonumber \\
&\ & \times 
\Bigg\{\, \left[ \frac{\partial^{2}}{\partial x^{2}} \right.  
\left. \left( \frac{1}{x} T_{q}(x,A) H_1(x) \right) \right]
\cdot \frac{2q_{T}^{2}}{(x's+u-Q^2)^{2}} 
\nonumber \\
&\ & \ \ \ \ 
+\left[ \frac{\partial}{\partial x}
\left( \frac{1}{x} T_{q}(x,A) H_1(x) \right) \right] 
\cdot \frac{2(Q^2-u)}{x's(x's+u-Q^2)} \, \Bigg\} \ ,
\label{e25}
\end{eqnarray}
where $x$ is given in Eq.~(\ref{ex}), and $\sigma^{DY}_0$ is the parton level 
Born Drell-Yan cross section
\begin{equation}
\sigma^{DY}_0=\frac{4\pi \alpha_{em}^2}{9Q^2} \, .  
\label{sdy0}
\end{equation}
In Eq.~(\ref{e25}), $H_1(x)$ is the partonic hard part  
\begin{equation}
H_1(x) = C_1 x \,\hat{H}_{1a} (x, x_1=0, x_2=0), 
\label{e26}
\end{equation}
and $\hat{H}_{1a}(x, x_1=0, x_2=0)$ is defined 
in Eq.~(\ref{e14}) with $k_T=0$. In Eq.~(\ref{e25}), we already include the
factor $(1/2\pi)(e^2/3Q^2)$ from the leptonic part of phase space (see 
Eq.~(\ref{e4b})).
 
After convoluting Eqs.~(\ref{e25}) with the corresponding parton 
distribution from the
beam, we obtain double scattering contribution to 
$hA\rightarrow l\bar{l}$ from the  first type soft-hard double 
scattering process 
\begin{eqnarray}
\frac{d\sigma_{1}^{(SH)}}{dQ^2 dq_T^2 dy}
&=& \sum_q \int dx' f_{\bar{q}/h}(x') 
\frac{d\hat{\sigma}_{1}^{(SH)}}{dQ^2 dq_T^2 dy} \, , 
\label{sha} 
\end{eqnarray}
where the parton level double scattering part 
$d\hat{\sigma}_{1}^{(SH)}/dQ^2 dq_T^2 dy$ is given in 
Eq.~(\ref{e25}).

Following the same derivations, we obtain contributions from second type and 
third type processes shown in Fig.~\ref{fig5} and Fig.~\ref{fig6}. 
For second type diagrams, 
the final result  for second type soft-hard 
process contributing to  
$hA\rightarrow l\bar{l}$ is
\begin{eqnarray}
\frac{d\sigma_{2}^{(SH)}}{dQ^2 dq_T^2 dy}
&=& \sum_q \int dx' f_{q/h}(x') 
\frac{d\hat{\sigma}_{2}^{(SH)}}{dQ^2 dq_T^2 dy} \, ,
\label{shb} 
\end{eqnarray} 
where the parton level double scattering part 
$d\hat{\sigma}_{2}^{(SH)}/dQ^2 dq_T^2 dy$ is:
\begin{eqnarray}
\frac{d\hat{\sigma}^{(SH)}_{2}}{dQ^2dq^2_Tdy} &=&
e_q^2\, \sigma_0^{DY} \, (12\pi \alpha_s^2)\, 
\frac{1}{2x's}\, \frac{1}{x's+u-Q^2}\,\nonumber \\
&\ & \times
\Bigg\{\, \left[ \frac{\partial^{2}}{\partial x^{2}}  
\left( \frac{1}{x} T_{g}(x,A) H_{2}(x) \right) \right]
\cdot \frac{2q_{T}^{2}}{(x's+u-Q^2)^{2}} 
\nonumber \\
&\ & \ \ 
+\left[ \frac{\partial}{\partial x}
\left( \frac{1}{x} T_{g}(x,A) H_{2}(x) \right) \right] 
\cdot \frac{2(Q^2-u)}{x's(x's+u-Q^2)} \, \Bigg\} \ .
\label{e29}
\end{eqnarray}
In Eq.~(\ref{e29}), the four-gluon matrix element $T_g$ is defined 
as \cite{LQS} 
\begin{eqnarray}
T_{g}(x,A) &=& 
 \int \frac{dy^{-}}{2\pi}e^{ixp^{+}y^{-}}
 \int \frac{dy_1^{-}dy_{2}^{-}}{2\pi} 
      \theta(y^{-}-y_{1}^{-})\theta(-y_2^{-}) \nonumber \\
&\ & \times \frac{1}{xp^{+}} 
     \langle p_{A}|F^{\sigma+}(y_{2}^{-}) F_{\alpha}^{\ +}(0)
                   F^{+\alpha}(y^{-})F_{\ \sigma}^{+}(y_1^{-})
     |p_{A}\rangle\ .
\label{Tg}
\end{eqnarray}
The partonic hard part $H_{2}$ in Eq.~(\ref{e29}) is defined as 
\begin{equation}
H_{2} = C_{2} \, \hat{H}_{2a}(x,x_1=0,x_{2}=0)\ ,
\label{e30}
\end{equation}
where $C_{2}$ is the overall color factor for second type diagrams. 
$\hat{H}_{2a}$ is given by the diagram shown in
Fig.~\ref{fig5}, and defined as
\begin{equation}
\hat{H}_{2a}=\frac{1}{4}\, 
{\rm Tr}\left[\gamma\cdot(x'p'+x_1p+k_T)
         R_{2a}^{\beta\nu}\, \gamma\cdot p_3\, 
         L_{2a}^{\alpha\mu}\right]
\left(-g_{\alpha\beta}\right)\left(-g_{\mu\nu}\right)\ ,
\label{e31}
\end{equation}
where $p_3=x'p'+(x+x_1)p-q$ is the momentum carried by the quark going
to final state.  

Similarly, for the third type soft-hard process, as 
sketched in Fig.~\ref{fig6}, 
the final result for third type soft-hard process  contributing to 
$hA\rightarrow l\bar{l}$ is
\begin{eqnarray}
\frac{d\sigma_{3}^{(SH)}}{dQ^2 dq_T^2 dy}
&=& \sum_q \int dx' f_{g/h}(x') 
\frac{d\hat{\sigma}_{3}^{(SH)}}{dQ^2 dq_T^2 dy} \, .
\label{shc} 
\end{eqnarray}
where the parton level double scattering part 
$d\hat{\sigma}_{3}^{(SH)}/dQ^2 dq_T^2 dy$ is:
\begin{eqnarray}
\frac{d\hat{\sigma}^{(SH)}_{3}}{dQ^2dq^2_Tdy} &=&
e_q^2\, \sigma_0^{DY} \, (12\pi \alpha_s^2)\,
\frac{1}{2x's}\, \frac{1}{x's+u-Q^2}\,\nonumber \\
&\ & \times 
\Bigg\{\, \left[ \frac{\partial^{2}}{\partial x^{2}}  
\left( \frac{1}{x} T_{q}(x,A) H_{3}(x) \right) \right]
\cdot \frac{2q_{T}^{2}}{(x's+u-Q^2)^{2}} 
\nonumber \\
& \ & \ \ 
+\left[ \frac{\partial}{\partial x}
\left( \frac{1}{x} T_{q}(x,A) H_{3}(x) \right) \right] 
\cdot \frac{2(Q^2-u)}{x's(x's+u-Q^2)} \, \Bigg\} \ .
\label{e32}
\end{eqnarray}
In Eq.~(\ref{e32}), the partonic hard part $H_{3}$ is defined as 
\begin{equation}
H_{3} = C_{3} x \, \hat{H}_{3a}(x,x_1=0,x_{2}=0)\ ,
\label{e33}
\end{equation}
where $C_{3}$ is the overall color factor for the third type process, 
and $\hat{H}_{3a}$ is given by the diagram shown in
Fig.~\ref{fig6}, and defined as
\begin{equation}
\hat{H}_{3a}=\frac{1}{4}\, 
{\rm Tr}\left[\gamma\cdot p\,
         R_{3a}^{\beta\nu}\, \gamma\cdot p_3\, 
         L_{3a}^{\alpha\mu}\right]
\left(-g_{\alpha\beta}\right)\left(-g_{\mu\nu}\right)\ ,
\label{e34}
\end{equation}
where $p_3$ is the same as that defined following Eq.~(\ref{e31}).

The partonic short-distance hard parts, defined in Eqs.~(\ref{e26}),
(\ref{e30}) and (\ref{e33}), can be easily evaluated by calculating
the corresponding Feynman diagrams. Our results are 
\begin{mathletters}
\label{e35}
\begin{eqnarray}
H_1(x) &=& \frac{2}{27}  \left( \frac{x-x_a}{x} \hat{\sigma}_1 \right.
             +\frac{x}{x-x_a} \hat{\sigma}_2
             \left. +\frac{2x_a}{(x-x_a)} \hat{\sigma}_3  \right)
\label{e35a} \\
H_{2}(x) &=& \frac{1}{36} \, \left( \hat{\sigma}_3+ \hat{\sigma}_4 \right.
          -\frac{2(x-x_a)}{x^2} \,\frac{Q^2}{x's}
          \left. \hat{\sigma}_1  \right)
\label{e35b} \\
H_{3}(x) & =& \frac{1}{16}  \left( \frac{x}{x-x_a} \hat{\sigma}_5 \right.
             +\frac{x-x_a}{x} \hat{\sigma}_6
             \left. -\frac{2x_a}{(x-x_a)} \hat{\sigma}_4  \right) \ ,
\label{e35c} 
\end{eqnarray}
\end{mathletters}
and
\begin{mathletters}
\label{hsigma}
\begin{eqnarray}
\hat{\sigma}_1 &=& \frac{Q^2-u}{x's+u-Q^2} 
\label{e36a} \\
\hat{\sigma}_2 &=& \frac{x's+u-Q^2}{Q^2-u}
\label{e36b} \\
\hat{\sigma}_3 &=& \frac{x's}{x's+u-Q^2}
\label{e36c} \\
\hat{\sigma}_4 &=& \frac{x's+u-Q^2}{x's}
\label{e36d} \\
\hat{\sigma}_5 &=& \frac{x's}{Q^2-u}
\label{e36e} \\
\hat{\sigma}_6 &=& \frac{Q^2-u}{x's}
\label{e36f} 
\end{eqnarray}
\end{mathletters}
In Eq.~(\ref{e35}),  $x_a$ is defined as
\begin{equation}
x_a=\frac{Q^2}{Q^2-u}.
\label{xa}
\end{equation}
In a special case, $Q^2=0$, our result for soft-hard process 
corresponds to direct photon production \cite{GQ}. In Eqs.~({\ref{e25}), 
(\ref{e29}) and (\ref{e32}), the only unknown quantities are the 
twist-4 parton distributions defined in Eqs.~(\ref{e24}) and (\ref{Tg}). 
They were first introduced by the authors of Ref. \cite{LQS}. These 
twist-4 parton distributions are universal functions, and as fundamental 
as the usual twist-2 parton distributions. They can be measured in one 
process and applied to another process. More discussions about these 
twist-4 distributions will be given in Sec. VI.
 
\section{double-hard scattering contribution}
\label{sec:dh}

In this section, we derive the contribution from double-hard processes. 
The calculations for double-hard  processes are simpler than soft-hard
processes. In soft-hard double scattering case, we use covariant gauge. In 
order to extract the contribution
beyond leading-twist in the covariant gauge, we have to keep $k_T$ for 
the soft gluon in collinear
expansion   and expand 
the hard-part at $k_T=0$. 
For double-hard processes, since the leading behavior is twist-four
contribution, we do not need to do such $k_T$ expansion. In addition, 
only the symmetric diagrams contribute to double-hard scattering. 
At lowest order, there are three types of diagrams contributing to 
double-hard subprocesses, as shown in Fig.~\ref{fig8}, Fig.~\ref{fig9} 
and Fig.~\ref{fig10}. 
In the following subsections, we 
give the derivations for one subprocess and provide the results for 
other subprocesses.

\subsection{Separation of matrix elements and partonic hard part}
\label{subsec:dh1}

Consider subprocess shown in Fig.~\ref{fig8}. Similar to soft-hard case, 
there are three independent four-momenta linking the partonic part and the
corresponding two-quark-two-gluon matrix element. After collinear expansion,  
the parton-nucleus double scattering contribution for subprocess shown 
in Fig.~\ref{fig8}
can be written as:
\begin{equation}
\frac{d\sigma^{(DH)}_{qA\rightarrow\gamma ^*}}{dQ^2dq^2_Tdy} 
= \frac{1}{2x's}\,\int dx\, dx_{1}\, dx_{2}\,
 \overline{T}^D(x,x_{1},x_{2},p)\,
\overline{H}^{DH}(x'p',x,x_{1},x_{2},p,q)\ .
\label{e37}
\end{equation}
The superscript ``$DH$'' denotes double-hard scattering. The double-hard  
two-quark-two-gluon matrix element, $\overline{T}^D$, is defined as: 
\begin{eqnarray}
\overline{T}^D(x,x_{1},x_{2},p) 
&=& \int \frac{p^+dy^{-}}{2\pi} 
     \frac{p^+dy_{1}^{-}}{2\pi} \frac{p^+dy_{2}^{-}}{2\pi}  
 e^{ixp^{+}y^{-}}\, e^{ix_{1}p^{+}y_{1}^{-}}\,
            e^{-i(x_{1}-x_{2})p^{+}y_{2}^{-}}\,
\nonumber \\
&\ & \times 
\frac{1}{2}\frac{1}{p^+}\langle p_{A} | 
            A^{\sigma}(y_{2}^{-},0_{T})\, \bar{\psi}_q(0)\, \gamma^+ \,
            \psi_q(y^{-})\, A^{\rho}(y_{1}^{-}) | p_{A}\rangle\ 
             (-g_{\sigma \rho}^{\perp}).
\label{e38}
\end{eqnarray}
$(-g_{\sigma \rho}^{\perp})$ represents physically polarized gluons. 
In Eq.~(\ref{e37}), the partonic part $\overline{H}^{DH}$ is given by the
diagram shown in Fig.~\ref{fig8}, with gluon lines contracted with
$(-g^{\rho \sigma})/2$, quark lines from the target traced with
$(\gamma\cdot p)/2$, and quark lines from the beam traced with 
$(\gamma\cdot (x'p'))/2$. 

In order to perform the integration of momentum fractions, we rewrite 
the double scattering contribution defined in 
Eq.~(\ref{e38}) as
\begin{eqnarray}
\frac{d\hat{\sigma}^{(DH)}_{qA\rightarrow\gamma^*}}{dQdq^2_Tdy} 
&=& \frac{1}{2x's}\, 
    \int \frac{dy^{-}}{2\pi}\, \frac{p^+dy_{1}^{-}}{2\pi}\,  
      \frac{p^+dy_{2}^{-}}{2\pi}\,
    H^{DH}(y^-,y_1^-,y^-_2,p,q)\ .
\nonumber \\ 
& & \times
\frac{1}{2}\, \langle p_{A} | 
            A^{\sigma}(y_{2}^{-},0_{T})\, \bar{\psi}_q(0)\, \gamma^+ \,
            \psi_q(y^{-})\, A^{\rho}(y_{1}^{-}) | p_{A}\rangle\ 
             (-g_{\sigma \rho}^{\perp}).
\label{e39}
\end{eqnarray}
In Eq.~(\ref{e39}), the modified partonic part $H^{DH}$ is defined as
\begin{eqnarray}
H^{DH}(y^-,y_1^-,y_2^-,p,q) 
&=& \int dx\, dx_{1}\, dx_{2}\, 
     e^{ixp^{+}y^{-}}\, e^{ix_{1}p^{+}y_1^{-}}\,
     e^{-i(x_{1}-x_{2})p^{+}y_{2}^{-}}  \nonumber \\
&\ & \times \overline{H}^{DH}(x'p',x,x_1,x_{2},p,q)\ ,
\label{e40}
\end{eqnarray}
All the integrals of the momentum fractions can now be done
explicitly without knowing the details of the matrix elements.

For diagram shown in Fig.~\ref{fig8}, the final state
photon-gluon two particle phase space can be written as
\begin{equation}
\Gamma^{(2)}=\frac{1}{16\pi^2}\, \frac{1}{x's+u-Q^2}\,
\delta\left(x+x_1+\frac{\hat{t}}{x's+u-Q^2}\right)\ .
\label{e41}
\end{equation}
Again, we neglect $dQ^2dq_T^2dy$ in the $\Gamma^{(2)}$ because of 
the definition of the differential cross section. 
Using Eq.~(\ref{e41}), the contribution to $\overline{H}^{DH}$ from 
the diagram
shown in Fig.~\ref{fig8} can be expressed as
\begin{eqnarray}
\overline{H}^{DH}_{1} &=& 
\frac{1}{16\pi^2}\, \frac{1}{x's+u-Q^2}\,
\delta\left(x+x_1+\frac{\hat{t}}{x's+u-Q^2}\right)
       \nonumber \\
&\times & \left(\frac{1}{3}\right)\, \frac{1}{(Q^2-u)}\,
\frac{1}{x-x_{a}-i\epsilon}\,
\frac{1}{x+x_2-x_a+i\epsilon}\,\hat{H}^{DH}_{1}(x,x_1,x_{2})\ .
\label{e42}
\end{eqnarray}
where the subscript ``1'' stands for first type double hard scattering 
subprocess shown in
Fig.~\ref{fig8}. $x_a$ is defined in Eq.~(\ref{xa}). The function 
$\hat{H}^{DH}_1$ in Eq.~(\ref{e42}) can be represented by the diagram 
shown in Fig.~\ref{fig11}a. It is given by
\begin{equation}
\hat{H}^{DH}_{1}=\frac{1}{4}\,
{\rm Tr} \left[ \gamma\cdot x'p' R_{1}^{\beta\sigma} 
        \gamma\cdot (q-x_1 p)\, L_{1}^{\alpha\rho} \right]
\left(-g_{\alpha\beta}\right)\left(-g_{\sigma\rho}\right)\ ,
\label{e43}
\end{equation}
where $R_{1}^{\beta\sigma}$ and $L_{1}^{\alpha\rho}$ are the right and
left blob, respectively, as shown in Fig.~\ref{fig8}.  These blobs
include all possible tree Feynman diagrams with the external partons
shown in the figure. The factor $1/3$ in Eq.~(\ref{e42}) comes from averaging 
over the color of the quark which annihilates to produce $\gamma^*$. 
We write it out explicitly here so that the remaining color factor will be just
the color factor for 
$\hat{H}^{DH}_1$, which is represented by the diagram in 
Fig.~\ref{fig11}a.

Substituting Eq.~(\ref{e43}) into Eq.~(\ref{e40}), 
we perform    
$dx_1$  integration  by using the $\delta$-function,
and $dx$, $dx_2$ by contour integration. We then have 
\begin{eqnarray}
H^{DH}_{1} &=&
\frac{1}{12}\, \frac{1}{x's+u-Q^2}\, \frac{1}{(Q^2-u)}\,
e^{i x_a p^{+}y^{-}}\, 
e^{i x_b p^{+}(y_1^{-}-y^{-}_2)} \nonumber \\
&\ & \times
\theta(-y^{-}_2)\, \theta(y^{-}-y_1^{-})\, 
\hat{H}^{DH}_{1}(x,x_1,x_2)\ ,
\label{e44}
\end{eqnarray}
where the $\theta$-functions result from the contour integrations,
and the momentum fractions for the function $\hat{H}^D_{1}$ are
\begin{mathletters}
\label{e45}
\begin{eqnarray}
x &=&x_a=\frac{Q^2}{Q^2-u} 
\label{e45a} \\
x_1 &=&x_b=-\frac{\hat{t}}{x's+u-Q^2}-x_a
\label{e45b} \\
x_2 &=& 0
\label{e45c}
\end{eqnarray}
\end{mathletters}
Substituting Eq.~(\ref{e44}) into Eq.~(\ref{e39}), we obtain   
\begin{eqnarray}
\frac{d\hat{\sigma}^{(DH)}_{qA\rightarrow\gamma^*}}{dQ^2dq^2_Tdy} 
&=& \left(\frac{8\pi^2\alpha_s^2\alpha_{em}}{3}\right)\,
   \frac{1}{2x's}\, 
    \int \frac{dy^{-}}{2\pi}\, 
       \frac{(p^+)^2dy_{1}^{-}dy_2^-}{2\pi}\,  
        e^{ix_ap^{+}y^{-}} e^{ix_bp^+(y_1^--y_2^-)}
\nonumber \\
&\ & \times
\frac{1}{2}\, \frac{1}{(x_bp^+)^2} 
\langle p_{A} | F_{\sigma}^{\ +}(y_{2}^{-},0_{T})\, \bar{\psi}_q(0)\, 
\gamma^+ \, \psi_q(y^{-})\, F^{+\sigma}(y_{1}^{-}) | p_{A}\rangle 
\nonumber \\ 
& & \times     \theta(-y^{-}_2)\, \theta(y^{-}-y_1^{-})\, 
\frac{1}{x's+u-Q^2}\, \frac{1}{(Q^2-u)}
 \hat{H}^{DH}(x,x_1,x_2)\ .
\label{e46}
\end{eqnarray}
In Eq.~(\ref{e46}), we already convert the field $A^{\sigma}$ into field
strength $F^{+\sigma}$ by partial integration. 

Define the double hard matrix
element \cite{LQS}
\begin{eqnarray}
T_{qg}^D(x_a,x_b,A) &=& \frac{1}{x_bp^+}
 \int \frac{dy_1^{-}}{2\pi}
 \int \frac{dy^{-}}{4\pi} \int p^+dy_{2}^{-} 
      \theta(y^{-}-y_1^{-})\theta(-y_{2}^{-}) 
        e^{ix_ap^{+}y^{-}} e^{ix_bp^+(y_1^--y_2^-)}\nonumber \\
&\ & \times 
     \langle p_{A}|F_{\sigma}^{\ +}(y_{2}^{-})\bar{\psi}_{q}(0)
                  \gamma^{+}\psi_{q}(y^{-})F^{+\sigma}(y_1^{-})
     |p_{A}\rangle\ .
\label{e47} 
\end{eqnarray}
Substituting Eq.~(\ref{e47}) into Eq.~(\ref{e46}), we obtain the 
parton level cross section for first type double hard process 
\begin{eqnarray}
\frac{d\hat{\sigma}^{(DH)}_{1}}{dQ^2dq^2_Tdy} 
&=& e^2_q\, \sigma^{DY}_0 \, (4\pi\alpha_s^2) 
 \frac{1}{x's+u-Q^2}\, \frac{1}{(Q^2-u)}\, \nonumber \\
&\ & \times \frac{1}{2x'x_b s}\,  
 T_{qg}^D(x_a,x_b,A) \hat{H}^{DH}_{1} \, ,
\label{e48}
\end{eqnarray}
with $\sigma_0^{DY}$ the parton level Drell-Yan Born cross section defined 
in Eq.~(\ref{sdy0}), and partonic hard part $\hat{H}_1^{DH}$ given  
in Eq.~(\ref{e43}). 
In deriving Eq.~(\ref{e48}), we already included the
factor $(1/2\pi)(e^2/3Q^2)$ from the leptonic part of phase space.
 
Comparing Eq.~(\ref{e48}) with Eq.~(\ref{e25}), we see that the 
contribution from double-hard process does not have the derivative 
terms, while the contribution from soft-hard process depends on the 
derivatives of the distributions and hard part. 
The derivatives of $x$ in Eq.~(\ref{e25}) actually result from 
the derivatives with respect to $k_T$ in Eq.~(\ref{e23}). As we pointed out in
Sec. III-A, for soft-hard 
process, the leading term when $k_T=0$ is the leading twist eikonal
contribution and does not correspond to physical double scattering. The 
lowest order high-twist contribution comes from the second order derivative
term in $k_T$ expansion. For double-hard process, since momentum fraction 
of both partons are finite, the leading contribution is already a twist-four 
contribution.        

\subsection{Final results for double-hard processes}
\label{subsec:dh2}

Convoluting Eq.~(\ref{e48}) with the
corresponding parton distribution from the beam hadron, we obtain 
contribution to $hA\rightarrow l\bar{l}$ from first type 
double-hard process shown in Fig.\ref{fig8}:  
\begin{eqnarray}
\frac{d\sigma_{1}^{(DH)}}{dQ^2 dq_T^2 dy}
&=& \sum_q \int dx' f_{\bar{q}/h}(x') 
\frac{d\hat{\sigma}_{1}^{(DH)}}{dQ^2 dq_T^2 dy} \ , 
\label{dha} 
\end{eqnarray}
with $d\hat{\sigma}_{1}^{(DH)}/dQ^2 dq_T^2 dy$ given in Eq.~(\ref{e48}). 

Following similar derivations, we obtain contributions from second and 
third  type double-hard processes shown in Fig.~\ref{fig9} and  
Fig.~\ref{fig10}. For contribution from second type 
double-hard process shown in Fig.~\ref{fig9}, we have
\begin{eqnarray}
\frac{d\sigma_{2}^{(DH)}}{dQ^2 dq_T^2 dy}
&=& \sum_q \int dx' f_{g/h}(x') 
\frac{d\hat{\sigma}_{2}^{(DH)}}{dQ^2 dq_T^2 dy} \ .
\label{dhb} 
\end{eqnarray}
The parton level double scattering part 
$d\hat{\sigma}_{2}^{(DH)}/dQ^2 dq_T^2 dy$ is given by 
\begin{eqnarray}
\frac{d\hat{\sigma}^{(DH)}_{2}}{dQ^2dq^2_Tdy} 
&=& e^2_q\, \sigma_0^{DY} \, (4\pi\alpha_s^2)
 \frac{1}{x's+u-Q^2}\, \frac{1}{(Q^2-u)}\, \nonumber \\ 
&\ & \times \frac{1}{2x'x_bs}\,
 T_{qg}^D(x_a,x_b,A) \hat{H}^{DH}_{2}.
\label{e49}
\end{eqnarray}
$\sigma_0^{DY}$ is defined by Eq.~(\ref{sdy0}). 
In Eq.~(\ref{e49}), the partonic hard part $\hat{H}^{(DH)}_{2}$ is 
given by 
\begin{equation}
\hat{H}^{(DH)}_{2}=\frac{1}{4}\,
{\rm Tr}\left[\gamma\cdot p_3 R_{2}^{\beta\sigma} \right.
       \left. \gamma\cdot (q-x_a p)\, L_{2}^{\alpha\rho}\right]
\left(-g_{\alpha\beta}\right)\left(-g_{\sigma\rho}\right)\ .
\label{e50}
\end{equation}
The diagram representing $\hat{H}^{(DH)}_{2}$ is shown in Fig.~\ref{fig11}b.

For the third type double-hard process, shown in 
Fig.~\ref{fig10}, we have
\begin{eqnarray}
\frac{d\sigma_{3}^{(DH)}}{dQ^2 dq_T^2 dy}
&=& \sum_q \int dx' f_{g/h}(x') 
\frac{d\hat{\sigma}_{3}^{(DH)}}{dQ^2 dq_T^2 dy} \, .
\label{dhc} 
\end{eqnarray}
And $d\hat{\sigma}_{3}^{(DH)}/dQ^2 dq_T^2 dy$ is given by  
\begin{eqnarray}
\frac{d\hat{\sigma}^{(DH)}_{3}}{dQ^2dq^2_Tdy} 
&=& e^2_q\, \sigma_0^{DY} (4\pi\alpha_s^2) 
\frac{1}{x's+u-Q^2}\, \frac{1}{(Q^2-u)}\,
\nonumber \\
&\ & \times  
 \frac{1}{2x'x_bs} T_{qq}^D(x_a,x_b,A) \hat{H}^{(DH)}_{3}.
\label{e51}
\end{eqnarray}
In Eq.~(\ref{e51}), the partonic hard part 
$\hat{H}^{(DH)}_{3}$ is given by 
\begin{equation}
\hat{H}^{(DH)}_{3}=\frac{1}{4}\, 
{\rm Tr}\left[\gamma\cdot x_b p R_{3}^{\beta\sigma} \right.
       \left. \gamma\cdot (q-x_a p)\, L_{3}^{\alpha\rho}\right]
\left(-g_{\alpha\beta}\right)\left(-g_{\sigma\rho}\right)\ .
\label{e52}
\end{equation}
The diagram representing $\hat{H}^{(DH)}_{3}$ is shown 
in Fig.~\ref{fig11}c.
In Eq.~(\ref{e51}), the four-quark matrix element is defined as\cite{LQS}
\begin{eqnarray}
T_{qq}^D(x_a,x_b,A) &=& 
 \int \frac{dy^{-}}{4\pi}
 \int \frac{dy_1^{-}}{4\pi} \int p^+dy_{2}^{-}
      \theta(y^{-}-y_1^{-})\theta(-y_{2}^{-}) 
        e^{ix_ap^{+}y^{-}} e^{ix_bp^+(y_1^--y_2^-)}\nonumber \\
&\ & \times 
     \langle p_{A}|\bar{\psi}_q(y_{2}^{-})
                   \gamma^+\psi_q(y_1^-) \bar{\psi}_{q}(0)
                  \gamma^{+}\psi_{q}(y^{-})
     |p_{A}\rangle\ .
\label{Tqq} 
\end{eqnarray}

The partonic short-distance hard parts $\hat{H}^{(DH)}_1$, 
$\hat{H}^{(DH)}_2$ and
$\hat{H}^{(DH)}_3$, defined in 
Eqs.~(\ref{e43}), (\ref{e50}) and (\ref{e52}), can be evaluated by 
calculating corresponding Feynman
diagrams in Fig.~\ref{fig11}. The results are:
\begin{mathletters}
\label{e53}
\begin{eqnarray}
\hat{H}^{(DH)}_1 &=&   \frac{4}{9} 
             \left( \frac{x's}{Q^2-u} \right.
             + \left. \frac{Q^2-u}{x's} \right)
             + \frac{(x's)^2+(Q^2-u)^2}{(x's+u-Q^2)^2}\, ;
\label{e53a} \\
\hat{H}^{(DH)}_{2} &=& \frac{1}{6} 
             \left( \frac{x's+u-Q^2}{Q^2-u} \right.
             + \left. \frac{Q^2-u}{x's+u-Q^2} \right)
         - \frac{3}{8} \frac{(x's+u-Q^2)^2+(Q^2-u)^2}{(x's)^2}\, ;
\label{e53b} \\
\hat{H}^{(DH)}_{3} &=& \frac{4}{9} 
             \left( \frac{x's}{x's+u-Q^2} \right.
             + \left. \frac{x's+u-Q^2}{x's} \right)
     +  \frac{(x's)^2+(x's+u-Q^2)^2}{(Q^2-u)^2}\, .
\label{e53c} 
\end{eqnarray}
\end{mathletters}
Eq.~(\ref{e53}) already includes the color factor for each individual Feynman
diagram. 

In above derivations, the only unknown quantities are double-hard matrix
elements $T_{qg}^D$ and $T_{qq}^D$, defined in Eqs.~(\ref{e47}) and 
(\ref{Tqq}). These double-hard matrix elements are universal functions.
They cannot be calculated perturbatively, but can be measured in one 
process and tested in another process. Theoretical predictions for double
scattering can also be tested once these matrix elements are measured. 

\section{relation between soft-hard and double hard scattering}
\label{sec:relation}

As discussed in Sec.~II, the contributions from soft-hard processes 
and double-hard
processes correspond to the contributions from  different residues 
when we integrate over the parton momenta using contour
integration.
Let us now look at the relationship 
between the two kinds of double scattering process. 
As an example, we exam
the process shown in Fig.~\ref{fig2}. The full amplitude for the double
scattering has the general form:
\begin{equation}
M \sim \int dx_1 \frac{1}{x_1-x_{1a}+i\epsilon}
\frac{1}{x_1-x_{1b}+i\epsilon} F(k_T^2,x_1,x),
\label{sec5-e1}  
\end{equation}
where $x_1$ is the parton momentum fraction which needs to be 
integrated, and $x$ is the 
other parton momentum fraction. The sum of these two parton momentum 
fractions is fixed by the kinematics, such as $x'p'$, $p_3$ and $q$. 
In Eq.~(\ref{sec5-e1}), the pole $x_1-x_{1a}$ corresponds to 
the first, and $x_1-x_{1b}$ 
corresponds to the second propagator, as shown in the
figure.  The poles are given by  
\begin{mathletters}
\label{sec5-e2}
\begin{eqnarray}
x_{1a} &=& \frac{k_T^2}{x's} \ ,
\label{x1a}
\\
x_{1b} &=& 
\frac{k_T^2+2p_3\cdot k_T+2p_3 \cdot x'p'}{x's-2p_3\cdot p}\ ,  
\label{x1b}
\end{eqnarray}
\end{mathletters}
and they are both in the same half of the complex plane in the region 
of interests ($x's-2p_3\cdot p > 0$).
In Eq.~(\ref{sec5-e1}), the function $F(k_T^2, x_1,x)$ is a non-vanishing 
and smooth function when $x_1=x_{1a}$ and $x_1=x_{1b}$. It is proportional 
to the parton fields of momentum at $x_1p$ and $xp$. 

When $q_T\neq 0$ and $k_T\rightarrow 0$, $x_{1a}\rightarrow 0$ while 
$x_{1b}$ is finite. 
Using contour integration to carry out $dx_1$-integral 
 in Eq.~(\ref{sec5-e1}), we have
\begin{eqnarray}
M &\sim & \frac{F(k_T^2,x_{1a},x_{tot}-x_{1a})}{x_{1a}-x_{1b}} -
\frac{F(k_T^2,x_{1b},x_{tot}-x_{1b})}{x_{1a}-x_{1b}} \nonumber \\
&\sim & M_{soft-hard} -M_{double-hard} \ ,
\label{sec5-e3}
\end{eqnarray}
where $x_{tot}$ is the sum of the total mometum fraction from the 
target, and is a function of $x'$, $p_3$ and $q$. 
Note that in Eq.~(\ref{sec5-e3}), the amplitude of these two 
terms have the opposite sign.
The square of these two terms gives the leading  high-twist 
contribution of double
scattering. Square of the first term corresponds to the soft-hard double 
scattering case, and square of the
second term corresponds to the double-hard double scattering case as 
we derived in the last two sections. 
As we will discuss later, the interference term is small when $q_T$ is 
large. 
   
When $q_T=0$, $p_{3_{T}}=k_T$, and the two poles given in Eq.~(\ref{sec5-e2}) 
become
\begin{mathletters}
\label{sec5-e4}
\begin{eqnarray}
x_{1a} &=&  \frac{k_T^2}{x's}\ ,  
\\
x_{1b} &=& \frac{x'\sqrt{s}/(2p_{3z})-1}
{x's-2\sqrt{s} p_{3z}-\sqrt{s}\,k_T^2/(2p_{3z})} \, k_T^2
\approx \frac{k_T^2}{2\sqrt{s}\, p_{3z}} \ .
\end{eqnarray}
\end{mathletters}
As  $k_T \rightarrow 0$, both $x_{1a}$ and $x_{1b}$ vanish.
To understand the region where $q_T \sim 0$ and $k_T \sim 0$, 
all terms proportional to $x_1$ (or $x_{1a}$, or $x_{1b}$) should 
not be included in $F(k_T^2,x_1,x=x_{tot}-x_1)$ 
for the leading pole contribution. Therefore, for 
the leading pole contribution in the region of $q_T \sim 0$ 
and $k_T \sim 0$, $F \sim F(x_{tot})$ and 
\begin{equation}
M_{soft-hard} \sim \frac{F(x_{tot})}{x_{1a}-x_{1b}} 
\sim M_{double-hard} \ . 
\label{sec5-lim}
\end{equation}
Consequently, $M \sim M_{soft-hard}-M_{double-hard}$ gives no leading
contribution to double scattering when $q_T=0$ \cite{QS-RHIC}. 

The double-hard process corresponds to the classical double scattering
picture. The contributions of double-hard processes given in Eqs.~(\ref{e48}), 
(\ref{e49}) and (\ref{e51}) have the form of 
\begin{equation}
\sigma \sim \sigma_1 \cdot \sigma_2 \cdot \frac{1}{Q^2} \ , 
\label{sec5-e5}
\end{equation}
where $\sigma_1$ and $\sigma_2$ are the partonic cross section for 
first and second 
scatterings respectively. The soft-hard process does not have the 
corresponding classical double scattering picture, and hence the cross section 
does not have the form of Eq.~(\ref{sec5-e5}).  The 
fact that the scattering amplitude for double-hard process and soft-hard 
process cancel each other when $q_T=0$ is a pure quantum effect. 

From Eq.~(\ref{sec5-e3}), we can see that the interference term of 
soft-hard
double scattering and the double-hard double scattering will  have 
a negative 
sign. That will give ``suppression'' effect. 
The interference is proportional to 
$F^{*}(k_T^2,x_{1a},x_{tot}-x_{1a})F(k_T^2,x_{1b},x_{tot}-x_{1b})$ 
or its complex conjugate. As $k_T \rightarrow 0$, 
$F^{*}(k_T^2,x_{1a},x_{tot}-x_{1a})F(k_T^2,x_{1b},x_{tot}-x_{1b})
\rightarrow F^{*}(0,0,x_{tot})F(0,x_{1b},x_{tot}-x_{1b})$. 
If $x_{1b} \neq 0$, overlap in phase space for $F^{*}$ and $F$ is 
clearly small because of the difference in parton momenta, while 
full overlap takes place when 
$x_{1b} \rightarrow 0$. 
Therefore, in large $q_T$ region, we
only consider the contribution from soft-hard and double-hard process
respectively, and neglect the interference between them. 
However,  in smaller $q_T$ 
region, one of the two hard partons in the double-hard amplitude 
will become softer,  and may have much bigger phase space to overlap 
with the soft-hard amplitude. Consequently, 
the interference term may become important and give large  
suppression effect. In order to obtain correct $q_T$ dependence for smaller 
$q_T$ region, we need to consider the interference between soft-hard and 
double-hard scattering as well as doing resummation of powers of  
$\alpha_s ln^2 (Q^2/q_T^2)$ \cite{GQS}.

\section{Numerical results and discussions}
\label{sec6}

In this section, we present our numerical results for the nuclear enhancement
in Drell-Yan transverse momentum distribution. 
We numerically evaluate the ratio of differential cross section 
to single
scattering contribution $R$ defined in Eq.~(\ref{e3}) by
using our analytical results presented in Sec. III and
Sec. IV.

The ratio $R$ defined in Eq.~(\ref{e3})
depends on contributions from both single scattering and double
scattering.  All these contributions depend on the non-perturbative
parton distributions or multi-parton correlation functions.  In
deriving following numerical results, the Set-1 pion distributions of
Ref.~\cite{Pion} are used for pion beams; and the CTEQ3L parton
distributions of Ref.~\cite{CTEQ3} are used for free nucleons.  The
twist-4 multi-parton correlation functions defined in Eqs.~(\ref{e24}), 
(\ref{Tg}), (\ref{e47}) and (\ref{Tqq}) 
have not been well-measured yet.  By comparing the definitions of
the soft-hard correlation functions $T_q$ and $T_g$  with 
the normal twist-2 parton
distributions \cite{CS}, authors of Ref.~\cite{LQS} proposed
following approximate expressions for the twist-4 soft-hard 
correlation functions $T_q$ and $T_g$:
\begin{equation}
T_{i}(x,A)=\lambda^{2}\, A^{1/3}\, f_{i/A}(x,A)
\label{e55}
\end{equation}
where $i=q,\bar{q}$, and $g$.  The $f_{i/A}$ are the effective twist-2
parton distributions in nuclei, and the factor $A^{1/3}$ is
proportional to the size of nucleus.  The constant $\lambda^2$ 
has dimensions of [energy]$^2$ due to the 
dimension difference between twist-4 and twist-2 matrix elements.  
The value of
$\lambda^2$ should be determined by experimental measurement. It was 
estimated in Ref.~\cite{LQS2} by using the
\underline{measured} nuclear enhancement of the momentum imbalance of
two jets in photon-nucleus collisions \cite{E683,E609}, and was found 
to be order of 
\begin{equation}
\lambda^{2} \sim 0.05 - 0.1 \mbox{GeV}^{2}\ .
\label{e56}
\end{equation}
This value is not too far away from the naive expectation from the
dimensional analysis, $\lambda^2 \sim \Lambda_{\rm QCD}^2$.  The theoretical 
predictions for the nuclear dependence in direct photon production base on this
value was consistent with the experiment \cite{GQ}. In our
calculation below, we use $\lambda^2 = 0.05$~GeV$^2$.  

  From the definition of the correlation functions 
in Eqs.~(\ref{e24}) and (\ref{Tg}), 
the lack
of oscillation factors for both $y^-_1$ and $y^-_2$ integrals can in 
principle give nuclear enhancement proportional to $A^{2/3}$.  
The $A^{1/3}$ dependence is a result of the assumption that the
positions of two field strengths (at $y^-_1$ and $y^-_2$, respectively)
are confined within one nucleon. 

For double-hard matrix elements $T_{qg}$ and $T_{qq}$ defined in 
Eqs.~(\ref{e47}) and (\ref{Tqq}), we adopt the model of Ref.\cite{MQ} and take 
\begin{mathletters}
\label{e57}
\begin{eqnarray}
T_{qg}(x_a,x_b,A)&=& \frac{C}{2\pi} A^{4/3} f_q(x_a) f_g(x_b) \ ;
\label{e57a} \\
T_{qq}(x_a,x_b,A)&=& \frac{C}{2\pi} A^{4/3} f_q(x_a) f_{\bar{q}}(x_b) \ ,
\label{e57b} 
\end{eqnarray}
\end{mathletters}
where $f_q, f_{\bar{q}}$  are normal twist-2 quark and anti-quark 
 distribution, and $f_g$ is a corresponding gluon distribution. 
$C=(0.35/8\pi r_0^2)$ GeV$^2$. $r_0$ is the value of nucleon radius, and  
$r_0 \approx 1.1-1.25$. Here we already convert the unit for $C$ to 
GeV$^2$. The factor $(1/2\pi)$ comes from the difference in the 
overall factor of the 
definition for matrix elements in this paper and in Ref.\cite{MQ}. 

In Eq.~(\ref{e55}), the effective nuclear parton distributions
$f_{i/A}$ should have the same operator definitions of the normal
parton distributions with free nucleon states replaced by the nuclear
states.  For a nucleus with $Z$ protons and atomic number $A$, we
define  
\begin{equation}
f_{i/A}(x,A)=A \left( \frac{N}{A}f_{i/N}(x)
+\frac{Z}{A}f_{i/P}(x) \right)\, R_i^{\rm EMC}(x,A)\ ,
\label{e58}
\end{equation}
where $f_{i/N}(x)$ and $f_{i/P}(x)$ with $i=q,\bar{q},g$ are normal
parton distributions in a free neutron and proton, respectively; and
$N=A-Z$.  The factor $R_i^{\rm EMC}$ takes care of the EMC effect in
these effective nuclear parton distributions.  We adopted the $R^{\rm
EMC}$ from Ref.~\cite{BQV}, which fits the data well.  

In Fig.~\ref{fig12}, we show the ratio of double scattering and single
scattering versus $q_T$. The ratio is normalized by $A^{1/3}$. The overall
feature is that the ratio becomes smaller for larger $Q$ and $q_T$. This is 
the typical behavior for high twist contribution. It is suppressed by the
factor of $1/Q^2$ or $1/q_T^2$. $Q$ and $q_T$ are both large scales in the 
processes. On the other hand, we see that when 
$q_T$ becomes very large, the ratio starts to rise again. This is the effect
from the edge of phase space. The soft-hard double scattering contribution
depends on the derivatives of the matrix elements. At edge of the phase 
space, the derivative term become more important. When $x\rightarrow 1$, 
the parton distribution $f(x)$ has $(1-x)^{\alpha}$ behavior, with 
$\alpha > 1$. As $x\rightarrow 1$, $d\, f(x)/dx >> f(x)$. Therefore, the
double scattering contribution becomes more important at the edge of the 
phase space. 
The curve for $Q=11$ GeV rise more steeply than the curve for 
$Q=5$ GeV, because 
at high $Q$, we approaches edge of phase space faster 
as we increase $q_T$. 

Fig.~\ref{fig13} shows the ratios of  
the soft-hard and the double-hard contributions to single scattering
contribution. The ratio is normalized by $A^{1/3}$. We see that the size 
of soft-hard and double-hard processes are comparable. At higher 
$q_T$, the ratio for  soft-hard process increase, as explained above. 
The contribution 
from double-hard processes increase faster when $q_T$ decreases. 
The contribution from double-hard processes depends on the product of two
parton distributions as given in Eq.~(\ref{e57b}). The momentum fractions 
$x_a$ and $x_b$ are smaller than the momentum fraction for soft-hard
processes. Because of the steep increase in parton distribution for smaller 
momentum fraction, the contribution from double-hard processes increase 
faster as $q_T$ becomes smaller. 

Fig.~\ref{fig14} shows the comparison of double 
scattering contribution and single scattering
contributions. The single scattering contribution is normalized by $A$ and 
the double scattering contribution is normalized by $A^{4/3}$. 
Without extra $A^{1/3}$ factor, the double scattering contribution 
are small compare to the single scattering contribution. Perturbative
calculation for double scattering should be  reliable. Because of $A^{1/3}$ 
enhancement, the double scattering becomes important. As $x_F$ increases, 
both double scattering contribution and single scattering 
contribution decreases,
but double scattering contribution decreases slower than single 
scattering. We expect larger 
enhancement in larger $x_F$ region. 

In Fig.~\ref{fig15}, we show the nuclear dependence of the $q_T$ spectrum in 
large $q_T$ region for different nuclear target: C, Ca, Fe, and  W,  when 
$800$ GeV  
proton beam is used. The figure is plotted for R versus $q_T$. The quantity 
R is defined in Eq.~(\ref{e3}). It is the ratio of total differential cross 
section and the single scattering cross section. Because of the $A^{1/3}$
enhancement for double scattering contribution, the ratio 
is larger for larger $A$.

As we discussed in the last section, when $q_T$ is smaller, the interference
between the double hard scattering and the soft-hard scattering becomes
important. The interference will give suppression. We expect that the
double scattering contribution will decrease as $q_T$ decreases in smaller
$q_T$ region.
Fig.~\ref{fig16} shows our 
expected ratio of double scattering contribution to single scattering
contribution. The dotted curve and the curve on the right side of the 
dash line is obtained by 
using our results presented in this paper. The curve on the left side of 
the dash line is our expected ratio when $q_T$ becomes smaller. 
Since the integrated cross section $d\sigma / dQ^2$ has little 
A-dependence, and we have shown nuclear enhancement in large $q_T$ region in
this paper,  we expect that there will be suppression in small $q_T$. 
In small $q_T$ region that the interference between the soft-hard and 
double-hard processes becomes dominant contribution.  
From the figure, we see that the calculated ratio starts to rise steeply 
when $q_T$ approaches $3$ GeV and below. This is the sign where the large 
logarithms of $\alpha_s ln^2 (Q^2/q_T^2)$ and the interference 
become important.  

In summary, both soft-hard and double-hard double scattering
processes contribute to the nuclear dependence in Drell-Yan pair 
production.   The double-hard subprocess resembles the classical double
scattering picture, while the soft-hard subprocess and the 
cancellation of the leading double-hard and soft-hard contribution 
are pure quantum 
effect.  
Both types of double scattering processes give $A^{1/3}$ nuclear 
enhancement. However, the interference between these two types of process 
have the suppression effect. In large $q_T$ region, the interference 
can be neglected and we have $A^{1/3}$ nuclear enhancement for Drell-Yan 
$q_T$ spectrum. In smaller $q_T$ region, the interference term 
becomes more important 
and may give nuclear suppression. Calculation of nuclear dependence of 
Drell-Yan process in small $q_T$ region will be addressed in a future 
publication \cite{GQS}.
 
\section*{Acknowledgments}

I thank Jianwei Qiu and George Sterman for very helpful discussions. 
I also thank Xin-nian Wang for helpful communications about the reference 
on the experimental data.
This work was supported by the Director, Office of Energy Research,
Division of Nuclear Physics of the Office of High Energy and Nuclear Physics of
the U.S. Department of Energy under contract No. DE-FG02-93ER40764.


\begin{figure}
\caption{Diagram for single scattering:
a) Annihilation diagram; 
b) ``Compton'' diagram.} 
\label{fig1}
\end{figure}

\begin{figure}
\caption{A Graphical representation of double scattering contributions
in proton-nucleus collisions.}
\label{figa}
\end{figure}

\begin{figure}
\caption{Sample Feynman diagrams contributing to next-to-leading order
double scattering: (a) soft-hard process; (b) double-hard process. 
 The circles represent the poles when we integrate over the  
parton momentum.}
\label{fig2}
\end{figure}

\begin{figure}
\caption{First type soft-hard double scattering subprocesses: 
``Annihilation'' diagrams corresponding to the
two-quark-two-gluon matrix element; 
a) real diagram, b) and c) interference diagram.}
\label{fig3}
\end{figure}

\begin{figure}
\caption{Second type soft-hard double scattering subprocesses: 
the real ``Compton'' diagrams corresponding to the
four-gluon matrix element.}
\label{fig5}
\end{figure}

\begin{figure}
\caption{Third type soft-hard double scattering subprocesses: 
the real ``Compton'' diagrams 
corresponding to the two-quark-two-gluon matrix element.}
\label{fig6}
\end{figure}

\begin{figure}
\caption{Diagram corresponding to first type  
double-hard subprocess.}
\label{fig8}
\end{figure}

\begin{figure}
\caption{Diagram corresponding to second 
type double-hard subprocess}
\label{fig9}
\end{figure}

\begin{figure}
\caption{Diagram corresponding to third 
type double-hard subprocess}
\label{fig10}
\end{figure}

\begin{figure}
\caption{
Diagrams representing the partonic hard parts for double-hard 
processes: 
a) $\hat{H}_{1}^{DH}$ for double-hard process  
shown in Fig.~\protect\ref{fig8}; 
b) $\hat{H}_{2}^{DH}$ for double-hard process 
shown in Fig.~\protect\ref{fig9}; 
c) $\hat{H}_{3}^{DH}$ for double-hard process 
shown in Fig.~\protect\ref{fig10}. 
}
\label{fig11}
\end{figure}

\begin{figure}
\caption{The ratio of double scattering contribution and single scattering 
contribution normalized by $A^{1/3}$ for a $800GeV$ proton beam when 
$Q=5GeV$ and $11GeV$.}
\label{fig12}
\end{figure}

\begin{figure}
\caption{The ratio of soft-hard and double-hard contributions to
single scattering contribution. The ratio is  
normalized by $A^{1/3}$.}
\label{fig13}
\end{figure}

\begin{figure}
\caption{Comparison of double scattering and single 
scattering contributions.}
\label{fig14}
\end{figure}

\begin{figure}
\caption{A-dependence of $q_T$ distribution at large $q_T$ 
region for  nuclear 
targets C, Ca, Fe and W. The quantity $R$ is the ratio of 
total differential cross section and the single scattering contribution, 
as defined in Eq.~(\protect\ref{e3}).}
\label{fig15}
\end{figure}

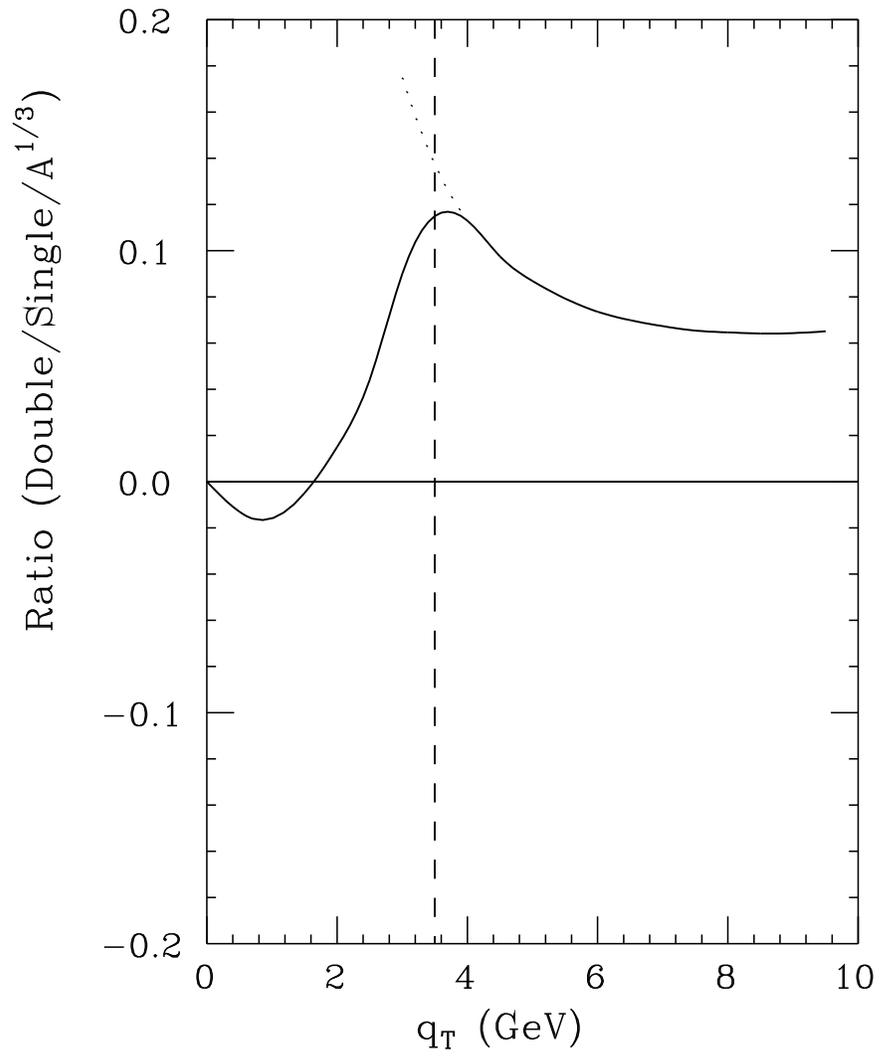
\begin{figure}
\caption{Expected ratio of double scattering to single scattering when 
$q_T$ is small.}
\label{fig16}
\end{figure}

\end{document}